\documentclass[useAMS,usenatbib,letters]{mn2e}
\usepackage[]{txfonts}
\usepackage{soul}
\usepackage{blindtext}
\usepackage{threeparttable}

\newcommand{\be}{\begin{equation}}
\newcommand{\ee}{\end{equation}}
\newcommand{\eqb}{\begin{eqnarray}}
\newcommand{\eqe}{\end{eqnarray}}

\newcommand{\pg}{\rm p\gamma}

\newcommand{\sth}{\sigma_{\rm T}}
\newcommand{\sgg}{\sigma_{\gamma \gamma}}

\newcommand{\zdiss}{z_{\rm diss}}
\newcommand{\mel}{m_{\rm e}}

\newcommand{\tpg}{t_{p\gamma}}
\newcommand{\gp}{\gamma^\prime_{\rm p}}
\newcommand{\Bmin}{B^\prime_{\min}}
\newcommand{\dmin}{\delta_{\min}}
\newcommand{\rbmin}{r^\prime_{\rm b,\min}}
\newcommand{\rb}{r^\prime_{\rm b}}

\newcommand{\bel}{\beta_{\rm e}}

\newcommand{\apr}{\alpha_{\rm p}}
\newcommand{\src}{3C~279\xspace}

\newcommand{\fermi}{{\it Fermi}-LAT \xspace}

\usepackage{graphicx}
\usepackage{epsfig} 
\usepackage{epstopdf}
\usepackage{longtable}
\usepackage{amssymb}
\usepackage{xcolor}
\usepackage{color}
\usepackage{lipsum}
\usepackage{textcomp}
\usepackage{threeparttable}
\usepackage[colorlinks=true,linkcolor=blue,citecolor=blue,
bookmarks=true,bookmarksopen=true,bookmarksnumbered=true,draft=false]{hyperref}
\usepackage{enumerate}

\title[A hadronic minute-scale GeV flare from 3C 279?]
{A hadronic minute-scale GeV flare from quasar 3C 279?}

\author[Petropoulou, Nalewajko, Hayashida, Mastichiadis]
{M. Petropoulou$^{1}$\thanks{Einstein Fellow; E-mail: mpetropo@purdue.edu}, K. Nalewajko$^{2}$, M. Hayashida$^{3}$, A. Mastichiadis$^4$\\
$^{1}$Department of Physics and Astronomy, Purdue University, 525 Northwestern
Avenue, West Lafayette, IN 47907, USA \\
$^2$ Nicolaus Copernicus Astronomical Center, Polish Academy of Sciences ,ul. Bartycka 18, 00-716 Warszawa, Poland\\
$^3$ Institute for Cosmic-Ray Research, University of Tokyo, 5-1-5 Kashiwanoha, Kashiwa, Chiba, 277-8582, Japan\\
$^4$ Department of Physics, National and Kapodistrian University of Athens,  Panepistimiopolis, GR 15783 Zografos, Greece
}
\begin{document}
\date{Received / Accepted}
\pagerange{\pageref{firstpage}--\pageref{lastpage}} \pubyear{2016}

\maketitle

\label{firstpage}

\begin{abstract}
The flat spectrum radio quasar 3C 279 is a known $\gamma$-ray variable source that has recently exhibited
minute-scale variability at energies $>100$~MeV. One-zone leptonic models for blazar emission are severely constrained by the short timescale variability that implies a very compact emission region at a distance of hundreds of Schwarzschild radii from the central black hole. Here, we investigate a hadronic scenario where GeV $\gamma$-rays are produced via proton synchrotron radiation. We also take into account the effects of the hadronically initiated electromagnetic cascades (EMC). For a $\gamma$-ray emitting region in rough equipartition between particles and kG magnetic fields, located within the broad-line region (BLR), the development of EMC redistributes the $\gamma$-ray luminosity to softer energy bands and eventually leads to broad-band spectra that differ from the observed ones. Suppression of EMC and energy equipartition are still possible, if the $\gamma$-ray emitting region is located beyond the BLR, is fast moving with Doppler factor ($>70$), and contains strong magnetic fields ($>100$~G). Yet, 
these conditions cannot be easily met in {parsec}-scale jets, thus disfavouring a proton synchrotron origin of the \fermi flare.

\end{abstract} 
  
\begin{keywords}
galaxies: active: individual: 3C 279 -- gamma-rays: galaxies -- radiation mechanisms: non-thermal
\end{keywords}

\section{Introduction} 
\label{sec:intro}
Variability ranging from few hours to several weeks has been commonly observed at various energy bands of the blazar spectrum. What came as a surprise was the detection of short (few minutes) timescale variability at very high energy $\gamma$-rays (VHE, $E_{\gamma}>100$ GeV)  from several blazars, including BL Lac objects [Mrk 421 \citep{fortson12}, Mrk 501 \citep{albert07}, and PKS 2155-304 \citep{aharonian07}] and flat spectrum radio quasars (FSRQ) \citep[PKS 1222+216,][]{aleksic07}. The latest addition to the above list is the minute-scale flare detected on June 2015 at  GeV $\gamma$-rays with \fermi  \citep{ackermann_16} from FSRQ \src \citep[at redshift z=0.536,][]{lynds65}.

The {2015} \fermi flare was characterized by high apparent $\gamma$-ray luminosity ($L_{\gamma}\sim 10^{49}$ erg s$^{-1}$), short flux-doubling timescales ($\lesssim 5$~minutes), and high values of the Compton dominance parameter ($\kappa=100$). The observed minute-scale variability suggests a very compact emission region at a distance of hundreds of Schwarzschild radii from the central black hole. In leptonic external Compton models (ECS)  \citep{dermeretal92, sikoraetal94}, the observed high Compton dominance requires a strongly matter-dominated emitting region \citep[see also][]{asano_hayashida15}. Models invoking the presence of relativistic protons in the jet have been proposed as an  alternative to the ECS scenario. In these models, $\gamma$-rays may result from a hadronically initiated electromagnetic cascade (EMC) \citep{mannheim91} or from relativistic proton synchrotron radiation \citep{aharonian00, mueckeprotheroe01}. 
The efficiencies of both hadronic processes are expected to be enhanced at sub-parsec dissipation distances \citep[see][for 3C 454.3]{khangulyan13}.

In this letter, we investigate if a proton synchrotron origin of the GeV flare is possible. The leptohadronic synchrotron model (LHS) for blazars is often criticized for its energetic requirements \citep[e.g.][]{zdziarskiboettcher15}. Recently, \cite{petro_dermer16} -- henceforth PD16, showed that the LHS model of the VHE  blazar radiation can have sub-Eddington absolute jet powers \citep[][]{cerrutietal15}, whereas models of dominant $>100$ MeV radiation in FSRQ may still require excessive power \citep[for 3C 273, see][]{petro_dimi15}. We do not discuss a photomeson ($\pg$) origin of the flare here, as this would have, in principle, higher energetic requirements than the LHS model \citep[for Mrk 421, see][]{dpm14}, and would result in flatter X-ray-to-$\gamma$-ray spectra than observed; a detailed calculation will be presented elsewhere.

As a starting point, we seek for parameters that minimize the jet power  and provide a successful fit to the time-averaged flare's spectrum from Orbit D \citep{ackermann_16}. To do so, we use the analytical expressions presented in PD16 and the main observables of the flare (variability timescale $t_{\rm v}$, $\gamma$-ray luminosity $L_{\gamma}$, and photon energy $E_{\gamma}$).
 It is possible that the produced $\gamma$-rays do not escape but they are absorbed by the softer photons in the source initiating EMC cascades
which alter drastically the broad-band spectrum as they distribute the $\gamma$-ray energy to lower frequencies, thus destroying the attempted spectral fit.
To assess the role of EMC, we derive analytical expressions for the optical depth for internal photon-photon ($\gamma \gamma$) absorption and the efficiency for $\pg$ interactions. The parameter values derived in the first step are then used as an input to a numerical code that calculates the broad-band photon spectrum taking into account all relevant physical processes \citep{DMPR2012}. In our numerical investigation, we compare the broad-band photon spectra obtained with and without internal $\gamma \gamma$ absorption.
 If, for parameters that minimize the jet power, the absorption of  VHE radiation produced via $\pg$ interactions initiates an EMC with a spectrum that does not describe well the data, we search for other parameters that may suppress the development of cascades.
 \begin{table}
 \caption{Observables of the minute-scale GeV flare from \src used in our analysis. Parameters describing the broad line region (BLR) are also  listed.}
 \begin{threeparttable}
 \begin{tabular}{llc}
  \hline
  Parameter & Symbol & Value \\
  \hline
  Variability timescale   (s) & $t_{\rm v}$ & $450$\\ 
  $\gamma$-ray luminosity (erg s$^{-1}$)    & $L_{\gamma}$ & $10^{49}$\\
  $\gamma$-ray photon energy (GeV) &  $E_{\gamma}$  & $0.7$ \\
  IR-to-UV luminosity (erg s$^{-1}$)  & $L$ & $10^{47}$\\
  Compton dominance parameter & $\kappa$ & 100 \\
  BLR luminosity\tnote{\textdagger} \, (erg s$^{-1}$)  & $L_{\rm BLR}$ & $6\times 10^{44}$\\
  BLR radius\tnote{\textdaggerdbl} \, (cm)  & $R_{\rm BLR}$&  $2.5 \times 10^{17}$ \\
  \hline
 \end{tabular}
 \begin{tablenotes}
 \item[\textdagger]  This has been estimated using the observed broad emission lines \citep{celotti97}.
 \item[\textdaggerdbl] Estimated using the scaling relation $R_{\rm BLR}\approx10^{17} L_{\rm d, 45}^{1/2}$ \citep{ghisellini_tavecchio08}.
 \end{tablenotes} 
 \end{threeparttable}
\label{tab-1}
\end{table}
\vspace{-0.2in}
\section{Analytical treatment}
\label{sec:results}
The observables of the flare that enter our analysis are presented in Table~\ref{tab-1}. Following PD16, we derive the parameter values that minimize the 
jet power in the LHS scenario. These are summarized in Table~\ref{tab-2} (Case A). In the following, the subscript ``min'' is used to denote quantities at minimum power conditions. The absolute minimum jet power is $P_{\rm j, \min} \simeq 14  \dot{M} c^2$ \citep[see also Fig. 2][]{ghisellini14}, where $\dot{M}c^2\simeq 10\, L_{\rm d}$ for a geometrically thin and optically thick accretion disk with bolometric luminosity $L_{\rm d} \simeq 6\times 10^{45}$~erg s$^{-1}$ \citep{pian99, hayashida_15}. 
At minimum power conditions, $\Bmin \sim 2$~kG and the emitting region is in rough equipartition between magnetic fields and relativistic protons ($u^{\prime}_{\rm p,\min}\sim u^\prime_{\rm B,\min}\simeq 2\times 10^5$ erg~cm$^{-3}$). Henceforth, all primed quantities are measured in the rest frame of the emitting source, while unprimed quantities are measured in the observer's frame. 
%
\begin{table}
 \caption{Parameter values that minimize the jet power of 3C 279 in the LHS scenario for the minute-scale GeV flare (Case A).  Other values considered in \S \ref{sec:numerical} are also listed (Case B). All primed quantities are measured in the rest frame of the emitting source. The beaming factor -- defined  as $\psi\equiv 2\Gamma/\delta$ -- was set equal to 1.01.}
 \begin{tabular}{llll}
\hline
Symbol & Parameter & Case A & Case B  \\
\hline
$\delta$  & Doppler factor & 19.5 & 50 \\
$\Gamma$& Bulk Lorentz factor & 9.9 & 25\\ 
$B^\prime$  (kG) & Magnetic field strength  & $2.2$ & 0.8\\
$\rb$ (cm) & Source radius & $1.7\times 10^{14}$ & $4.4\times 10^{14}$\\
$P_{\rm j}$ (erg s$^{-1}$)& Absolute total jet power & $8.3\times 10^{47}$  & $10^{48}$  \\
$u^\prime_{\rm p}$ (erg cm$^{-3}$)& Proton energy density  & $2.6\times 10^5$  &  $7\times 10^3$  \\
$u^\prime_{\rm B}$ (erg cm$^{-3}$)& Magnetic energy density   & $2\times 10^5$   &  $2.6\times 10^4$  \\
$E^\prime_{\rm p, \max}$  (eV)& Max. proton energy  & $5.2\times 10^{16}$ &  $4.8\times 10^{16}$\\ 
$\zdiss$ (cm)& Dissipation distance & $1.7\times 10^{16}$ & $1.1\times 10^{17}$\\
\hline
 \end{tabular}
\label{tab-2}
\end{table}
In such strong magnetic fields, the energy of protons producing synchrotron photons at the peak energy of the $\gamma$-ray component is  $E^\prime_{\rm p, \max} \sim 50$~PeV. This is lower  than the energy of protons that can be confined in the source by a factor $\sim 2400$ \citep{hillas_84}. 
 Because of the short variability timescale and strong magnetic fields, the synchrotron cooling timescale of the highest energy protons in the source is comparable to the dynamical timescale ($t^\prime_{\rm syn} \sim 3.5 t^\prime_{\rm dyn}\simeq 2\times 10^4$~s).

In the following, we assume that the size of the emitting source is approximately equal to the size of the dissipation region. Depending on its opening angle $\theta$, the dissipation distance may lie from several up to hundreds of gravitational radii ($r_{\rm g}$)
away from the base of the jet. Assuming that $\theta\approx \theta_{\rm obs}$ and $\Gamma \theta_{\rm obs}=0.1$, we find $z_{\rm diss,\min}\simeq \rbmin/\theta \simeq 143 \, r_{\rm g} < R_{\rm BLR}$. For a smaller degree of collimation (e.g., $\psi = 2$) the dissipation distance would be placed at $28\, r_{\rm g}$ from the central black hole, while the mimimum jet power would be $\psi^2$ times higher.

\begin{figure*}
 \centering
\includegraphics[width=0.49\textwidth]{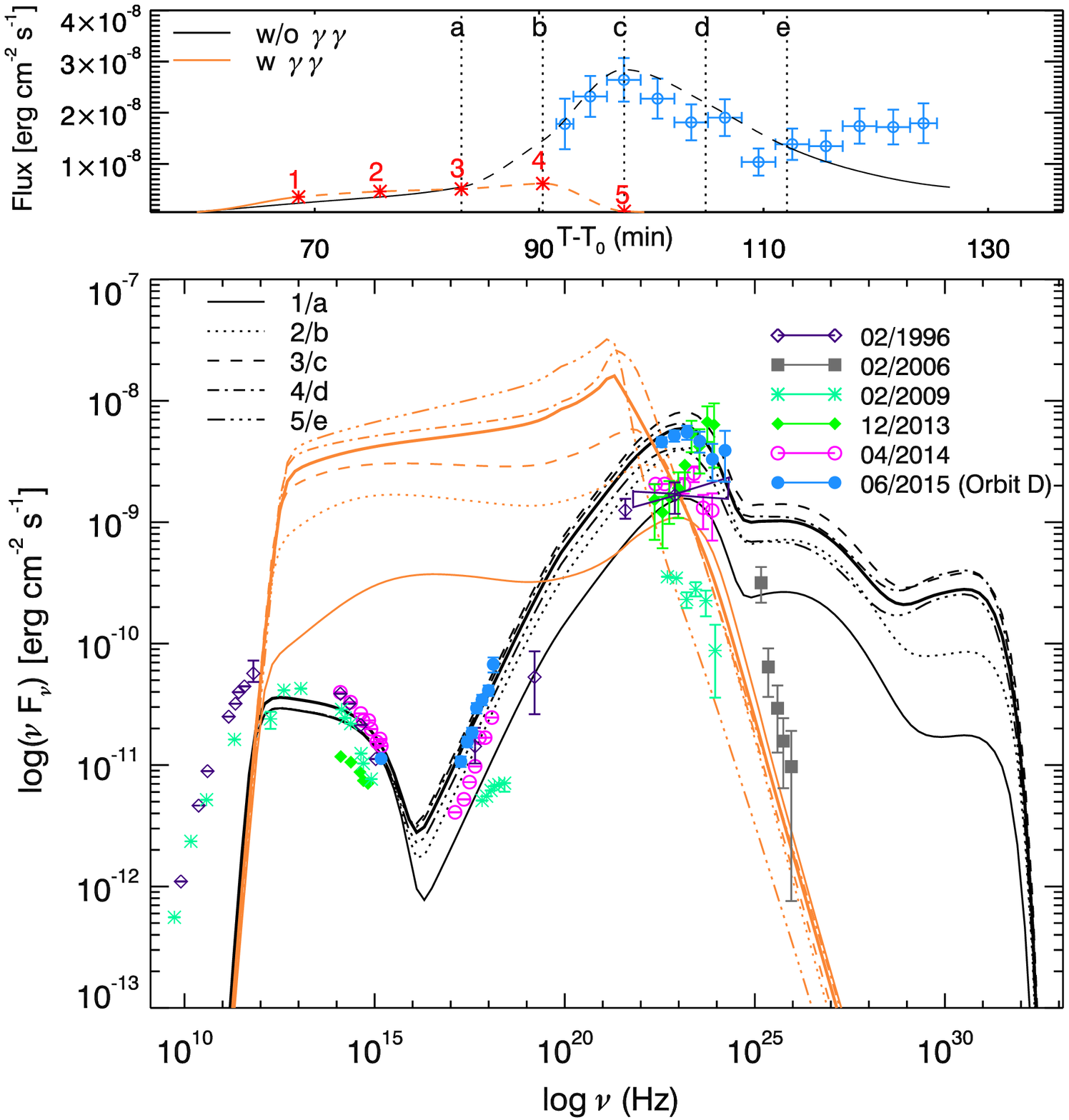}
\includegraphics[width=0.49\textwidth]{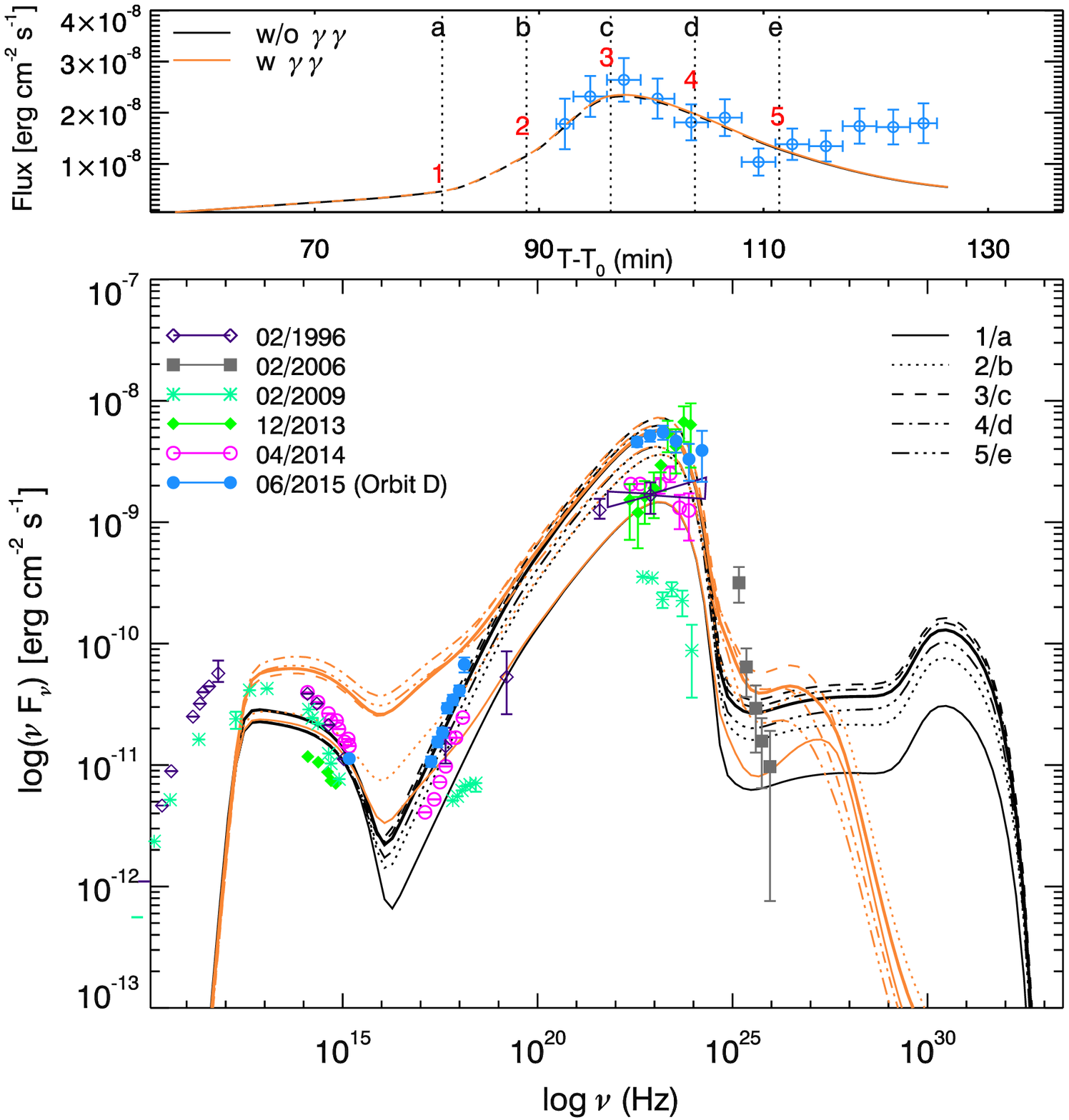}
\caption{The 3-minute binned \fermi light curve from Orbit D (top panels). Here, $T_0$ is defined as June 16 2015 02:00:00 (UT). For the conversion from photon to energy fluxes, a power-law spectrum with photon index $-2.1$ in the range $0.1-10$~GeV was assumed. The model light curves with (orange) and without {(black)}  $\gamma \gamma$ absorption are overplotted. The dashed part of the light curves indicates the period where the time-averaging of the model spectra has been performed. Vertical lines and asterisks denote the times of the model spectra presented in the bottom panel. SED of \src \ compiled using UV, X-ray and \fermi data of Orbit D (blue filled circles)  during the latest outburst on June 2015 (bottom panels). Multi-wavelength data of past flaring periods (adopted by \citet{ackermann_16}) are also shown with coloured symbols (see legend). SED snapshots during the outburst phase from Orbit D are presented with thin lines. The numbering and color coding of the curves
follows the notation in the top panel.  Thick lines show the model spectra averaged over a period of $\sim 30$~min (dashed lines on top panel). The displayed spectra do not take into account the attenuation on the extragalactic background light. {\sl Left panel:} The parameters are chosen as to fit the average spectrum of Orbit D and to minimize the jet power (Case A in Table~\ref{tab-2}). {\sl Right panel:} The parameters used (Case B) correspond to $\delta\simeq 2.5\dmin$, $B^\prime =0.4 \Bmin$, $P_{\rm j}=1.25\, P_{\rm j, min}$, and $\zdiss \simeq R_{\rm BLR}$. }
\label{fig:fig1}
\end{figure*}

We next provide estimates of the efficiency for $p\gamma$ interactions and the optical depth for internal $\gamma \gamma$ absorption using simplified expressions for the respective cross sections. We will take into account the radiation from the illumination of the BLR as well as the internal synchrotron radiation produced by the relativistic electron and proton distributions; photons from the high-energy hump of the spectrum have not been taken into account in previous works \citep[e.g.,][]{dermer07, sikora09}. For the purposes of the analytical treatment, we model the internal target photon field as a broken power law, i.e., $n^\prime_{\rm i}(\epsilon^\prime)= n^\prime_{\rm i,0}/\epsilon^{\prime 2} \left[\left(\epsilon^\prime/\epsilon^\prime_{\rm i}\right)^{\alpha_{\rm i}}H\left(1-\epsilon^\prime/\epsilon^\prime_{\rm i}\right) + \left(\epsilon^\prime/\epsilon^\prime_{\rm i}\right)^{\beta_{\rm i}} H\left(\epsilon^\prime/\epsilon^\prime_{\rm i}-1\right)\right],$
where $\epsilon^\prime$ is the photon energy in $\mel c^2$ units, $H(x)$ is the Heavyside function, $\alpha_{\rm i} > 0$, $\beta_{\rm i}\le 0$, $\epsilon^\prime_{\rm i}=\epsilon_{\rm i} (1+z)/\delta$, and  i=e (p) for electron (proton) synchrotron radiation. In particular, $\epsilon_{\rm  p}=E_{\gamma}/\mel c^2$ with $E_{\gamma}=0.7$~GeV (Table~\ref{tab-1}) and  $\epsilon_{\rm e}=8\times10^{-9} \nu_{12}$, where we introduced the notation $Q_{x}\equiv Q/10^x$ in cgs units. Our choice for the low peak energy is based on archival observations (see also Fig.~\ref{fig:fig1}). The normalization of the photon number density is given by $ n^\prime_{\rm i,0}= 3L_{\rm i}/(4\pi r^{\prime 2}_{\rm b} \delta^4 \mel c^3)$, where $\rb=c\delta t_{\rm v}/(1+z)$, $L_{\rm e}\equiv L$, and  $L_{\rm p}\equiv L_{\gamma}$ (Table~\ref{tab-1}). {The BLR energy density profile is assumed to be uniform up to a typical radius $R_{\rm BLR}$ \citep[e.g.,][]{ghisellini_tavecchio08, dermer14}. Its differential photon number density is 
approximated by} $n^\prime_{\rm BLR}(\epsilon^\prime) = 15u^\prime_{\rm BLR} (\mel c^2)^3 \epsilon^{\prime 2}  (e^{\epsilon^{\prime }\mel c^2/kT^\prime}-1)^{-1}/\left(\pi kT^\prime\right)^4$, where {$kT^\prime \simeq \Gamma E_{\rm BLR}$, $E_{\rm BLR}=10$~eV}, $u^\prime_{\rm BLR}\approx \Gamma^2 L_{\rm BLR}/ 4\pi c R_{\rm BLR}^2$, and {$L_{\rm BLR} \simeq 6\times 10^{44}$ erg s$^{-1}$ \citep{celotti97}.  {At $z_{\rm diss,min}$ the direct disk radiation can be neglected, since $u^\prime_{\rm d}/u^\prime_{\rm BLR}\simeq 0.08$. Here, we assumed that the size of the accretion disk that emits most of the bolometric radiation is $R_{\rm d}=10^{15}\sim 10 \, r_{\rm g}$ \citep{ghisellinimadau96, sikora09}. }}

The energy loss timescale due to $\pg$ interactions
is given by $ \tpg^{\prime -1}\left(\gp \right) =c/(2\gamma_{\rm p}^{\prime 2}) 
\int_{\bar{\epsilon}_{\rm th}}^{\infty} {\rm d}\bar{\epsilon} \sigma_{\pg}(\bar{\epsilon}) \kappa_{\pg}(\bar{\epsilon})\bar{\epsilon}
\int_{\bar{\epsilon}/ 2 \gp}^{\infty} {\rm d}\epsilon' n'(\epsilon')/\epsilon^{' 2}$ \citep{stecker68, sikoraetal87}, 
where $\gp$ is the proton's Lorentz factor,  bared quantities are measured in the proton's rest frame, $\bar{\epsilon}_{\rm th} \simeq 300$, $n'(\epsilon')$ is the number density of target photons. In the analytical calculations we use  $\sigma_{\pg} \approx 0.1\, {\rm mb} \, H(\bar{\epsilon} - \bar{\epsilon}_{\rm th})$, and $\kappa_{\pg}=0.2$. 
The exact cross section, inelasticity and pion multiplicity are used in the numerical calculations \citep{DMPR2012}. 
The efficiency for $\pg$ interactions, {defined as $\eta_{\pg}\equiv \rb /c t^\prime_{\pg}$,} on electron synchrotron photons ($\alpha_{\rm e}>1$ and $\beta_{\rm e}<0$) for $\gp \ll  10^{11}\delta_1/\epsilon_{\rm  e, -8}$ is 
\eqb
\eta_{\pg, \rm e} \simeq 1.4\times10^5 \frac{L_{47}t^{-1}_{v,2}\delta^{-4}_1\epsilon^{-1}_{\rm e,-8}}{ (1-\beta_{\rm e})(3-\beta_{\rm e})} 
                                                                    \left(\frac{10^{11}\delta_1}{\gp \epsilon_{\rm e,-8}}\right)^{-1+\beta_{\rm e}}. 
\label{eq:taupg_e}
\eqe
For parameters that minimize the jet power (Table~\ref{tab-2}) and $\beta_{\rm e}\sim -1/4$, we find $\eta_{\pg, \rm e} \ll 1$. The efficiency also depends strongly on the Doppler factor as $\eta_{\pg, \rm e} \propto \delta^{-5+\beta_{\rm e}}$. For $\pg$ interactions on proton synchrotron photons ($0<\alpha_{\rm p}<1$ and $\beta_{\rm p}<0$) and $\gp \gg \delta_1/\epsilon_{\rm  p, 3}$ we find 
\eqb 
\eta_{\pg, \rm p} = 1.3\times10^{-4}\frac{L_{49}t^{-1}_{v,2}\delta^{-4}_1 \epsilon^{-1}_{\rm p, 3}}{(1-\alpha_{\rm p})(3-\alpha_{\rm p})} 
                                                                   \left(\frac{2 \gp \epsilon_{\rm p, 3}}{\delta_1}\right)^{1-\alpha_{\rm p}}.
\label{eq:taupg_p}
\eqe
The efficiency increases with increasing proton energy as  $\gamma_{\rm p}^{\prime 1-\alpha_{\rm p}}$ and $\eta_{\pg, \rm p}\propto \delta^{-5+\alpha_{\rm p}}$, where $\alpha_{\rm p}$ corresponds to the X-ray-to-$\gamma$-ray photon index of the spectrum in the LHS model. For $\pg$ interactions with BLR photons from the Rayleigh-Jeans part of the spectrum, the efficiency is independent of the proton's energy and is given by
\eqb
\eta_{\pg, \rm BLR} \simeq 2\times10^{-5} \frac{t_{v,2} L_{\rm BLR, 45}\delta_1^2}{\epsilon_{\rm BLR,-5} R_{\rm BLR, 17}^2}, \gp \gg \frac{9\times 10^6}{\psi \delta_1  \epsilon_{\rm BLR,-5}}.
\label{eq:taupg_blr}
\eqe
In contrast to $\eta_{\pg, \rm e(p)}$ that have a strong inverse dependence on the Doppler factor, $\eta_{\pg, \rm BLR} \propto \delta^2$.
The optical depth for internal $\gamma \gamma$ absorption is $\tau_{\gamma \gamma}(\epsilon_1)=\rb \int {\rm d} \epsilon^\prime n^\prime(\epsilon^\prime)\sgg(\epsilon^\prime,\epsilon^\prime_1)$, where $\sgg(\epsilon^\prime,\epsilon_1^\prime)= 0.652 \, \sth\, H(\epsilon_1^\prime \epsilon^\prime-2)/(\epsilon_1^\prime \epsilon^\prime)$ \citep[e.g.,][]{coppiblandford90}. 
The respective optical depth due to electron synchrotron photons may be written as 
\eqb
\tau_{\gamma \gamma, \rm e} (\epsilon^\prime_1)\simeq  
                               \frac{6\times 10^{7+\bel}(1+z)^{1-\bel}}{2^{2-\bel}(2-\bel)}\frac{L_{47} \epsilon_{1,8}^{\prime 1-\bel}}{t_{\rm v,2}\delta_1^{5-\bel} \epsilon_{\rm e,-8}^{\bel}}, \, \epsilon_1^\prime \ll 10^{9} \frac{\delta_1}{\epsilon_{\rm e,-8}}.                                        
 \label{eq:taugg_e}
\eqe
In the above, we used $\alpha_{\rm e}>2$, since the electron synchrotron spectrum below its peak is usually synchrotron self-absorbed (see Fig.~\ref{fig:fig1}  below). Noting that $0<\apr <1$, the optical depth due to internal absorption on  proton synchrotron photons is given by
\eqb
\tau_{\gamma \gamma, \rm p}(\epsilon^\prime_1) \simeq \frac{(1+z)^{1-\apr}}{2^{2-\apr}(2-\apr)}\frac{L_{\gamma, 49} \epsilon_{1,-2}^{\prime 1-\apr}}{t_{v,2} \epsilon_{\rm p,3}^{\apr} \delta^{5-\apr}},  \, \epsilon_1^\prime \gg 10^{-2} \frac{\delta_1}{\epsilon_{\rm p, 3}}.
\label{eq:taugg_p}
\eqe
The optical depth increases for increasing photon energy as $\tau_{\gamma \gamma, \rm p}\propto \epsilon_1^{\prime 1-\apr}$, while it depends on the Doppler factor as $\delta^{-5+\apr}$.  Substitution of parameter values that minimize the jet power (see Table~\ref{tab-2}) and of the flare's observables (Table ~\ref{tab-1}) in eqs.~(\ref{eq:taugg_e}) and (\ref{eq:taugg_p}) shows that any internally produced radiation at $E \gtrsim 20$~GeV will be attenuated by the electron and proton synchrotron radiation fields. In fact, an EMC can be developed, since $\tau_{\gamma \gamma}\gg 1$ at multi-TeV energies. Depending on the luminosity of the multi-TeV photons and the exact value of $\tau_{\gamma \gamma}$, the  spectrum of photons originally considered  as targets will be modified. We expand on this issue in \S \ref{sec:numerical} with detailed numerical 
calculations. Finally, the optical depth due to the Rayleigh-Jeans part of the BLR photon energy spectrum may be written as
\eqb
\tau_{\gamma \gamma, \rm BLR} \simeq 0.6 \frac{\delta_1 t_{\rm v,2} L_{\rm BLR, 45}}{\epsilon_{1,5}^\prime \epsilon_{\rm BLR, -5}^2 R_{\rm BLR, 17}^2}, \, \epsilon_1^\prime \gg \frac{10^5}{\psi \delta_1 \epsilon_{\rm BLR, -5}}.
\label{eq:taugg_blr}
\eqe
In contrast to $\tau_{\gamma \gamma, \rm e(p)}$ which decrease with increasing $\delta$ (eqs.~(\ref{eq:taugg_e}) and (\ref{eq:taugg_p})) the respective optical depth  due to the BLR radiation field is $\propto \delta$  as long as the $\gamma$-ray emitting region lies within the BLR. 
\vspace{-0.2in}
\section{Numerical SED modelling}
\label{sec:numerical}
Using as a starting point the parameter values of Case A (Table~\ref{tab-2}), we perform time-dependent calculations of the broad-band spectrum aiming at   fitting the average spectrum of the outburst phase from orbit D \citep{ackermann_16}.
Simultaneous observational data were available in the UV, X-ray and $\gamma$-ray bands, but not in the radio-to-optical bands.
The latter are not crucial for our analysis, as our main conclusions remain unaffected by a different modelling of the low-energy component of the SED. 
\begin{table}
 \caption{Parameters of the particle distributions at injection as derived from the numerical modelling of the flare's time-averaged spectrum over the orbit~D for both cases shown in Fig.~\ref{fig:fig1}.}
 \begin{tabular}{ccc cc cc c}
\hline
$\gamma^\prime_{\rm p, \min}$ & $\gamma^\prime_{\rm p,  br}$ & $\gamma^\prime_{\rm p, \max}$ & $s_{\rm p, 1}$ & $s_{\rm p, 2}$ & $\gamma^\prime_{\rm e, \min}$ & $\gamma^\prime_{\rm e, \max}$ & $s_{\rm e}$\\
1 &  $10^6$ & $10^8$ & 1.6 & 2.1 & 1 & 160 & 2.1\\
\hline
 \end{tabular}
\label{tab-3}
\end{table}
Although  a detailed fit to the light curve during Orbit D lies beyond the scope of this letter, we do take into account the basic temporal properties of the  flare (i.e., duration, peak and average fluxes). {In an attempt to model the flare's broad-band spectrum  with the minimum number of changes in the model parameters, we imposed fluctuations on the proton energy density alone:}
$u^\prime_{\rm p}(\tau)={\tilde{u}^\prime_{\rm p}\left[1+4.75/(0.25+(\tau-5)^2)\right]}$, where $\tau\equiv t^\prime/t^\prime_{\rm dyn}=[0,10]$ and $\tilde{u}^\prime_{\rm p}={8}\times 10^4$  (800)  erg cm$^{-3}$ for Case A (Case B). The proton energy density averaged over ten dynamical timescales is $\langle u^\prime_{\rm p}\rangle={3}\times 10^5$ ($3\times 10^3$) erg cm$^{-3}$  for Case A (Case B), in agreement with our analytical estimates in Table~\ref{tab-2}. In both cases, we choose initial conditions that correspond to a low flux level of the source. In the numerical calculations we also lift some of the simplifying assumptions (e.g., monoenergetic proton distribution) used in PD16 and \S\ref{sec:results}. {In particular, the particle injection spectra are modelled as broken power laws, i.e, $n_{\rm i}^\prime(\gamma^\prime_{\rm i}) \propto \gamma_{\rm i}^{\prime -s_{\rm i,1}}S(\gamma_{\rm i}^{\prime}; 
\gamma_{\rm i,\min}^{\prime}, \gamma_{\rm i, br}^\prime) + \gamma_{\rm p}^{\prime -s_{\rm i,2}}S(\gp; \gamma_{\rm i, br}^\prime, \gamma^\prime_{\rm i, \max})$, where $S(y;y_1,y_2)$ is the unit boxcar function. The parameter values derived from the fits are presented in Table~\ref{tab-3}.}
Because of the derived kG magnetic fields, our numerical calculations also include muon and pion losses due to synchrotron radiation \citep{petro_dimi14}. 

Our results are shown in Fig.~\ref{fig:fig1}. In each panel, we present the 0.1-10 GeV light curves (top) 
and the broad-band photon spectra (bottom); 
for more details, see figure caption. The results obtained by neglecting (taking into account) internal $\gamma \gamma$ absorption are plotted, in both panels, with black  (light blue) lines.  For parameters that minimize the jet power (left panel), an EMC is being developed with a photon spectrum that cannot describe the observed one. Photons produced via photohadronic processes (at multi-TeV energies) are attenuated and their luminosity is transfered to lower (i.e., IR-to-X-rays) energies. The increase of the target photon number density in those softer bands increases, in turn, the efficiency for photohadronic interactions, thus leading to an increased production rate of VHE photons. This creates a feedback loop of processes that results in higher photon luminosities  and spectral shapes dramatically different than the target photon spectra before the initiation of the EMC \citep[see also][]{petromast12,petroGRB_14}. Because of the lower peak photon energy (few MeV) of the EMC spectrum, the model light 
curve in the 0.1-10 GeV lies far below the observed fluxes (left top panel in Fig.~\ref{fig:fig1}). We next seeked for parameters that would (i) provide a successful fit to the data from Orbit D, (ii) suppress the development of EMC, and (iii) lead to $P_{\rm j} \gtrsim P_{\rm j, \min}$ (Case B). The EMC is suppressed compared to Case A, but the internal absorption of VHE photons still leads to an excess below few keV. Because of the higher $\delta$, the $\pg$ photon production in Case B is driven by the BLR radiation field (\S\ref{sec:results}). 
\vspace{-0.3in}
\section{Discussion}
\label{sec:discuss}
{We have derived parameter values that minimize the jet power of \src during the minute-scale \fermi flare in the LHS model. For these conditions, however, an EMC is developed in the $\gamma$-ray emitting region due to the internal absorption of multi-TeV photons produced via photohadronic interactions. The most straightforward way for suppressing the development of an EMC is to decrease the number of relativistic pairs that are being injected in the emitting region by $\pg$ and $\gamma \gamma$ processes. This can be achieved in a faster moving emitting region because of the strong dependence of $\eta_{\pg, \rm e(p)}$ and $\tau_{\gamma \gamma, \rm e(p)}$ on $\delta$ (\S\ref{sec:results}). To ensure that $P_{\rm j} \gtrsim P_{\rm j, min}$, the magnetic field in the emitting region should be $B^\prime < \Bmin$ for $\delta > \dmin$, given that the main contributing terms to the total jet power scale as $P_{\rm j, B}\propto B^{\prime 2}\delta^4$ and $P_{\rm j, p}\propto B^{\prime -6} \delta^{-6}$. Taking 
also into account the radiation power, we find that $P_{\rm j}\simeq 1.25 P_{\rm j, min}$ for $\delta=2.5\dmin\simeq50$ and $B^{\prime}=0.8$~kG (Table~\ref{tab-2}). We have shown that a fit to the $\gamma$-ray spectrum of the flare is possible for this parameter set (Case B). However, our model spectra cannot describe well the observations below few keV due to the EMC emission. {The discrepancy between the data and the model could be reduced, if $u^\prime_{\rm BLR}$ was lower by a factor of ten than the adopted value. This would, in turn, suggest a more extended BLR ($R_{\rm BLR}>0.25$~pc) than the one predicted by the simple scaling relation of \citet{ghisellini_tavecchio08}. } 

{For a BLR with properties as those listed in Table~\ref{tab-1}, the effects of the EMC on the broad-band spectrum could be suppressed}, if the $\gamma$-ray emitting region was located outside the BLR \citep[see][for leptonic scenarios]{nalewajko14}. This would require $\delta \gg \sqrt{2  R_{\rm BLR} (1+z)\sqrt{\psi-1}/ct_{\rm v}\psi} \simeq 70$ for $\Gamma\theta_{\rm obs}=0.1$. The jet power could still be close to the minimum one provided that $B^\prime > 100$~G at distances $>0.1$~pc \citep[see also][]{bottacini16}. The main challenge is to explain then, the extremely high local energy densities \citep{2012MNRAS.425.2519N} associated with the strong magnetic fields. Such regions in {parsec}-scale jets would be severely over-pressured, and would not survive for more than a couple of dynamical time scales. Were the magnetic field strength lower (e.g., 10 G), the energetic requirements would be very high, i.e., $P_{\rm j}\gg 100 \dot{M}c^2$ \citep[e.g.][]{petro_dimi15}.

{It is noteworthy that a flat proton energy distribution ($s_{\rm p} \lesssim 2$) is required for explaining the high-energy hump of the SED \citep[see also][]{boettcherreimer13, cerrutietal15}, while $s_{\rm e} \sim 2$ (see Table~\ref{tab-3}). Magnetic reconnection in highly magnetized plasmas may account for flat particle energy spectra, as demonstrated  by particle-in-cell (PIC) simulations in electron-positron plasmas for $\sigma \gtrsim 10$ \citep{guo14, ss14, nalewajko15}. Similar behavior is expected in electron-proton plasmas for $\sigma \gg 1$ \citep{melzani14,guo16}, where $\sigma = B_0^2 / (4 \pi \rho_0 c^2)$ is the jet magnetization, $B_0$,  and $\rho_0 c^2$ are the un-reconnected magnetic field strength and plasma rest mass energy density, respectively. The latter can be estimated from the energy density at the blazar dissipation region, $u^\prime_{\rm j} \sim P_{\rm j}  \epsilon_{\rm r}^2/ (2 \pi c \Gamma^6 \theta^2 r_{\rm g}^2)$, where $\epsilon_{\rm r} = 0.1 \epsilon_{r, -1}$ is the 
reconnection rate \citep[see also][]{giannios13}. The jet's magnetic field strength far from the reconnection layer can be estimated from the magnetic field in the emission region as $B_0 \sim B^\prime/\sqrt{2}$ \citep{sgp16}. For typical parameters, we find that $B_0^2/(8\pi) \sim \rho_0 c^2 \sim u^\prime_{\rm j}/2$ or equivalently $\sigma \sim 1$. As PIC simulations of reconnection in moderately magnetized electron-proton plasmas are sparse, the properties of the accelerated particle distributions in this regime remain unclear (Dr. Sironi, private communication).}  


The minute-scale variability and high $\gamma$-ray luminosity characterizing the 2015 outburst from \src challenge both leptonic and leptohadronic one-zone models for the blazar emission. Allowing for relativistic motion of the emitting region in the rest frame of the jet, as proposed in models of magnetic reconnection \citep[e.g.,][]{gianniosetal09, nalewajko11, pgs16}, might resolve some of the issues that single-zone models encounter.

\vspace{-0.2in}
\section*{Acknowledgments}
We thank the anonymous referee for an insightful report and Prof. M. Sikora for useful comments on the manuscript. 
M.~P. acknowledges support from NASA through the Einstein Postdoctoral 
Fellowship grant number PF3~140113 awarded by the Chandra X-ray 
Center, which is operated by the Smithsonian Astrophysical Observatory
for NASA under contract NAS8-03060.
K.N. was supported by the Polish National Science Centre grant 2015/18/E/ST9/00580.
\vspace{-0.2in}
\bibliographystyle{mn2e} 
\bibliography{3c279}

\end{document}